# Dynamics of elastic, nonheavy spheres sedimenting in a rectangular duct

## Isabell Behrendt[1] and Clarissa Schönecker[1,2], †

[1]Technische Universität Kaiserslautern, D-67663 Kaiserslautern, Germany

[2]Max-Planck-Institut für Polymerforschung, D-55218 Mainz, Germany

Understanding of sedimentation dynamics of particles in bounded fluids is of crucial importance for a wide variety of processes. While there is a profound knowledge base regarding the sedimentation of rigid solid particles, the fundamental principles of sedimentation dynamics of elastic, nonheavy spheres in bounded fluids are not well understood. Therefore, we performed sedimentation of deformable, elastic solid spheres with $Re_P \ll 1$ in a model experiment. The spheres started from rest in a rectangular duct with a width of about 23 times the radius $R$ of the sphere. The particle dynamics of elastic spheres differed fundamentally from that of rigid spheres. Elastic effects were found to take place on comparatively large time scales, such that elastic spheres underwent four phases of sedimentation. Phases I and II, including transient acceleration and a short steady velocity plateau, are comparable with sedimentation of rigid spheres. From a characteristic onset position of about $10 \cdot R$, deformability effects kick in and a second acceleration appears (phase III). In the fourth phase, the deformable spheres reach the terminal sedimentation velocity. The softer the spheres are (in terms of Young's elastic modulus), the higher the terminal velocity is. In the present setup, a terminal velocity up to 9 % higher than the velocity for comparable rigid spheres was reached. By means of the obtained data, insights into the dynamics are given that could serve as basic approaches for modelling the dynamics of elastic spheres in bounded fluids.

† Email address for correspondence: schoenecker@mv.uni-kl.de

## 1. Introduction

Sedimentation due to gravity is a widely used process in civil and environmental engineering. In a variety of applications, like in wastewater treatment or in food industry, sedimentation is used specifically as for example separation method. (De 2017) In other applications, such as in the ink and paint industry, sedimentation of solid pigments is an undesirable effect that must be avoided. Profound knowledge and modelling of fluid-particle dynamics of sedimentation is therefore essential and it is the basis for the design of these processes.

The calculation and design of such processes is mostly based on the assumption of spherical, heavy and rigid particles. However, both in nature and in applications, the assumption of rigid and heavy particles is often not valid. For example, microplastics or microgels, such as silicone particles, are in many cases soft and deformable. Further, biological particles cannot be treated as rigid and heavy particles since, for example, fungal capsules or cell membranes are deformable. (Araújo et al. 2019)

Wastewater treatment plants have been designed in the past mostly using rigid particle sedimentation for design calculations. But in the process, increasing amounts of microplastics enter the wastewater. Shape factors and empirical formulations are used to account for the sedimentation velocity of microplastics, which obviously differs from that of other heavier particles. (Dietrich 1982) Furthermore, several polymers have a density insignificantly higher than that of water. The density ratio is therefore in the range of unity. These particles are called nonheavy particles. Studies show that a significant proportion of microplastics cannot be removed from wastewater in sewage treatment plants and end up in the environment. (Iyare et al. 2020)

The motion of biological particles such as bacteria or blood cells in flows is often described using, e.g. challenging soft capsule models or, e.g. flexible bond models in CFD-DEM simulations. (Barthès-Biesel 2016; Balachandran Nair et al. 2020) It is assumed, that elastic solid beads could also serve as good models for biological particles.(Villone and Maffettone 2019) Additionally, such soft particle models are of particular interest in the field of microfluids, like transportation of soft particles through rectangular micro ducts. Here, however, not only the link between material properties (deformability) and fluid dynamics plays a role, but also the influence of the surrounding walls. (Villone 2019)

While the literature regarding the dynamics of rigid particles is vast and consolidated, the research and knowledge regarding the dynamics of elastic particles is still incomplete in places. Therefore, it remains of great interest to the scientific community. (Villone and Maffettone 2019) Even comparatively simple particle dynamics, such as the sedimentation of elastic spheres in an unbounded fluid, is only described in basic theory or is based on numerical calculations. (Murata 1980; Takeuchi et al. 2010)

The aforementioned examples, in which the sedimentation of deformable particles factually plays a decisive role, show that basic research is necessary at this point. Therefore, we perform sedimentation experiments, which allow insights into the dynamics of elastic spheres at low Reynolds numbers in a rectangular duct. Trajectories and resulting dynamics are recorded on a macroscopic scale. With the data obtained, both the transient and the steady-state sedimentation of deformable particles under wall influence can be quantitatively determined and related to the deformability.

## 2. Theoretical Background

A rigid spherical particle with radius $R$, volume $V_P$ and density $\rho_P$ sediments due to gravitational acceleration $g$ with velocity $U_P$ in an unbounded fluid. The fluid density is $\rho$ and the fluid dynamic viscosity $\eta$. Furthermore, the particle starts from rest at $t = 0$ s and, initially, the fluid velocity field around the particle is $|U| = 0$ m·s⁻¹. In the stationary state the particle sediments in the Stokes regime with the particle Reynolds number $Re_P = 2U_P R \rho \eta^{-1} \ll 1$. The equation of motion of a nonheavy particle with low particle-to-fluid density ratio $\gamma = \rho_P/\rho$ can be described with the Basset-Boussinesq-Oseen (BBO) equation (according to Basset (Basset 1888), Boussinesq (Boussinesq 1903) and Oseen (Oseen 1927)). The BBO equation for the particle velocity $U_P$ in direction of the gravitational acceleration reads as (Clift, R., Grace, J., & Weber, M. 1978; Chang and Yen 1998)

$$\rho_P V_P \frac{dU_P}{dt} = (\rho_P - \rho)gV_P - 6\pi\eta R U_P - \frac{1}{2}\rho V_P \frac{dU_P}{dt} - 6R^2\sqrt{\rho\pi\eta}\int_0^t \frac{\dot{U}(s)ds}{\sqrt{t-s}} \quad . \tag{1}$$

The overdot denotes a time-derivate and $s$ is a dummy time variable for integration. The BBO equation results from the force balance around the particle with $F_I + F_g + F_{Buancy} + F_{St} + F_{AM} + F_B = 0$ (see figure 1 *(a)*). The term on the left side of equation (1) is the inertial force $F_I$. The inertial force is equaled by the gravitational force $F_g$ and the buoyancy $F_{Buancy}$. $F_{St}$ denotes the drag force resulting from Stokes´s law (Stokes 1851). The third term on the right is the added mass force $F_{AM}$ which arises because the acceleration of the particle requires acceleration of the fluid surrounding it. The fourth term is the Basset history force $F_B$. The Basset history force includes past acceleration and represents the effect of the diffusion of vorticity around the sphere due the temporal delay in boundary layer development. (Clift, R., Grace, J., & Weber, M. 1978; Michaelides 1997) When a rigid particle starts sedimenting from rest, it first undergoes a transient phase until it reaches a stationary, terminal sedimentation velelocity. The two acceleration forces $F_{AM}$ and $F_B$ lead to a more extended transient phase compared to heavy particles (see figure 1 *(b)*). In the steady state ($dU_P / dt = 0$), where the gravitational force and buoyancy balance the drag and the acceleration forces vanish, equation (1) results in the terminal Stokes velocity

$$U_{St} = \frac{2R^2 \Delta \rho g}{9\eta} \ .$$

(2)

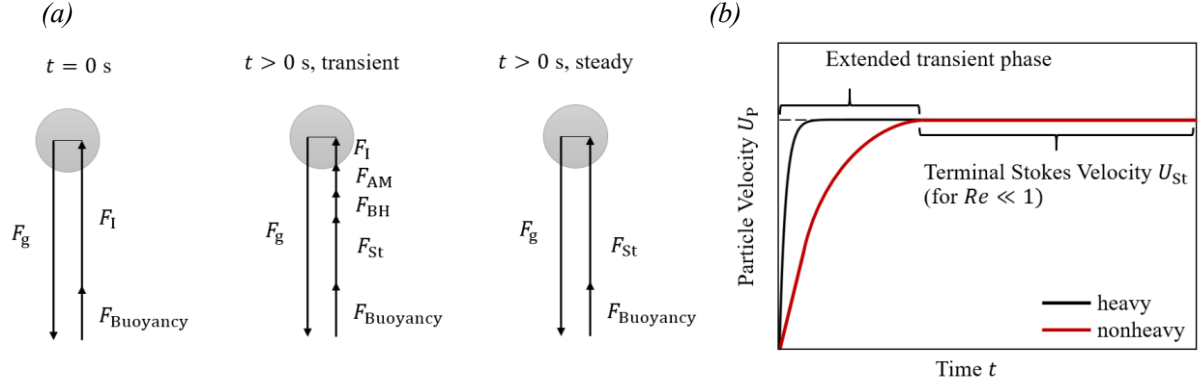

Figure 1: *(a)* Force balance around a spherical particle at different times during sedimentation; *(b)* Exemplary velocity curve of a sedimenting sphere with a high particle-to-fluid density ratio $\gamma$ (heavy) and a low particle-to-fluid density ratio $\gamma$ (nonheavy)

The previous equations are valid for particles sedimenting in an unbounded fluid. The influence of walls on the terminal velocity of a rigid sphere between parallel plane walls was first treated by Faxén. (Faxén 1922) A detailed overview of empirical equations and of mathematical methods for calculating the influence of walls on particle dynamics is given by Happel and Brenner. (Happel and Brenner 1983) In principle, the following statement can be made: if particles sediment in the presence of walls, drag exerted on the sphere is increased and thus the terminal velocity is reduced. While there are numerous empirical as well as theoretical approaches to calculate drag coefficient corrections for cylindrical pipes, drag corrections for confined spaces with plane walls are mostly based on computational studies. For example, in a computational study, Swan and Brady calculated the terminal velocity of a spherical particle sedimenting along a channel with two parallel walls in dependence of the wall distance. (Swan and Brady 2010) Since they assume the space between the two walls as infinite, the real terminal velocity is estimated to be even lower. These effects would be expected in confined spaces like in a rectangular duct. Hensley and Papavassiliou performed CFD simulations to calculate the drag exerted on a sphere moving in the center of a square microduct with width $W$. (Hensley and Papavassiliou 2014) For a spherical particle that moves with $Re_P = 0.1$ in a square microduct they proposed a correction of the Stokes´s drag. Using their correction factor, the particle velocity in a rectangular duct reduces to

$$U_{P,corr} = \frac{1}{1 + 5\dfrac{R}{W}} U_{St} \ .$$

(3)

In contrast to rigid solid particles, there is little knowledge available with respect to deformable, elastic solid particles. In 1980, Murata derived an expression for the terminal sedimentation velocity of an elastic sphere in an unbounded fluid. According to Murata´s calculations, when a sphere sediments in the low Reynolds number regime with $Re_P \ll 1$, it deforms into a prolate spheroid in direction of the gravitational acceleration. (Murata 1980) This deformation into a prolate spheroid would reduce the drag acting on the elastic body. Hence, the terminal sedimentation velocity increases. (Happel and Brenner 1983) By series expansion with the perturbation parameter $\alpha^2$ and in terms of specific constants $K_0$ and $K_1$, Murata derived the following expression for the terminal sedimentation velocity of an elastic sphere

$$U_{St,el} = \frac{2R^2 \Delta \rho g}{9\eta} \left(1 + 2\alpha^2 K_0 + \frac{1}{5}\alpha^2 K_1\right) = U_{St}(1 + \Delta_{el,M}) \ .$$

(4)

In this equation, the terminal velocity for rigid particles is extended by an elastic component of the terminal velocity $\Delta_{el,M}$. The perturbation parameter relates the viscous stresses to the elastic, mechanical stresses with $\alpha = U_{St} \eta G^{-1} R^{-1} \ll 1$. Since $\alpha$ is much smaller than unity, small deformations are assumed. The elastic constants $K_0$ and $K_1$ are calculated from the solid material properties ($K_{0,1} = f(G, \lambda, \rho_P, R, g)$). $G$ and $\lambda$ denote Lamé´s constants with $G = E[2(1 + \nu)]^{-1}$ and $\lambda = E\nu[(1 + \nu)(1 - 2\nu)]^{-1}$. $E$ represents Young´s elastic modulus and $\nu$ the Poisson´s ratio of the elastic solid material.

To the best of our knowledge, the theoretically derived velocity from equation (4) has not yet been verified by experimental data. Furthermore, it is not known whether the equation, or the increase in sedimentation velocity respectively, is analogous in the case of sedimentation in a bounded fluid. It is not yet known whether walls as in a rectangular duct have an additional influence not captured by Murata´s calculations.

## 3. Methods

### 3.1. Experimental setup

Sedimentation of rigid and elastic model particles was performed in a rectangular duct. As a vessel for the experiments, a glass container of 140 mm x 140 mm x 420 mm ($W$ x $D$ x $H$) was used, see figure 2 *(a)*. The container was filled with silicone oil which is a Newtonian fluid. The silicone oil had a nominal viscosity of 1000 cSt. The choice of the model particle size together with the choice of a high viscosity liquid ensured that $Re_P$ is always much smaller than 1. This condition allowed to investigate the same conditions as small, elastic microparticles would experience in aqueous liquids. The spheres were hold with a pipette under weak vacuum and then immersed in the liquid. The wall distance of the sphere center to all sides was about $d \approx 11R$ (see figure 2 *(b)*). After immersion, the vacuum was released. The sphere began to fall due to gravity. This ensured that the spheres began to sediment at an initial velocity of $U_{P,0} = 0 \text{ m} \cdot \text{s}^{-1}$ (starting from rest). High-resolution videos of the sedimentation experiments were recorded by a DSLR camera (Nikon® D7200 with Sigma® 50 mm f1.4 objective lens). Subsequently, the velocities were evaluated with a self-programmed image processing tool in MATLAB. During the experiments, the container with silicone oil was illuminated from the opposite side of the camera by a collimated light panel. This ensured a sharp contour of the model particle recorded by the camera sensor.

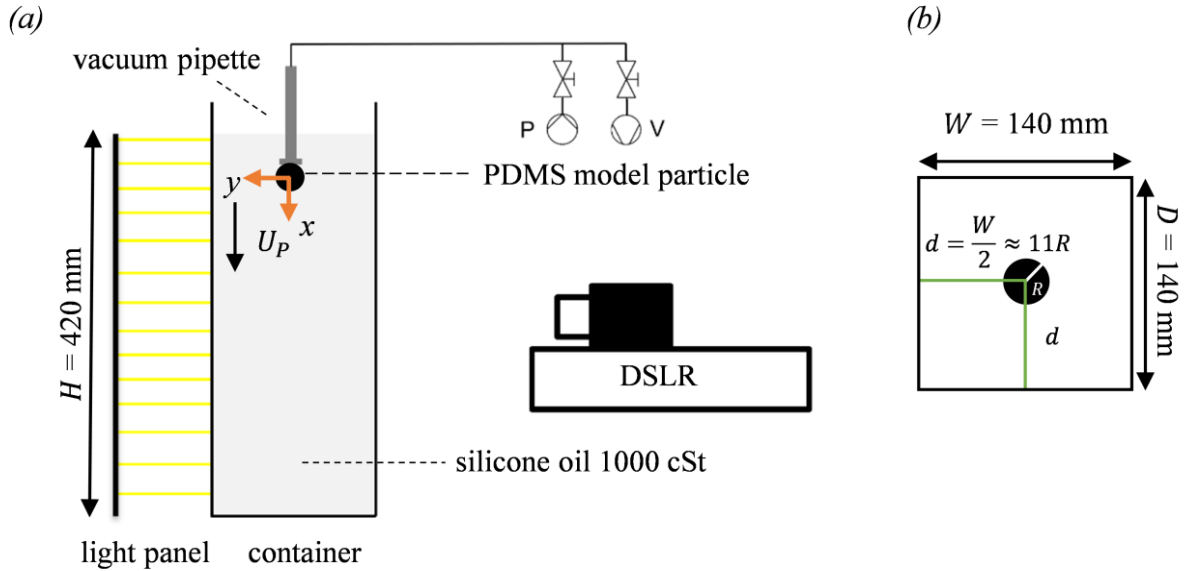

Figure 2: *(a)* Schematic experimental setup: spherical particles sediment in a rectangular container with square base in silicone oil with a nominal viscosity of 1000 cSt. The particles were immersed hanging on a vacuum pipette. By flooding the pipette with air, the spheres were released; *(b)* Top view and dimensions: The particles were placed in the center of the container so that the wall distance to all walls is $d \approx 11R$

### 3.2. Model particle fabrication

Elastic model particles were fabricated from polydimethylsiloxane (PDMS) by casting. To vary Young´s modulus of the particles, mixtures of two PDMS types were prepared in different mixing ratios. The fabrication procedure was based on the report by Palchesko et al. (Palchesko et al. 2012) First, base blends of commercially available PDMS elastomer Sylgard™ 184 (Dow Corning®) and PDMS Dielectric Gel Sylgard™ 527 (Dow Corning®) were prepared according to the manufacturer's specifications. The recommended mixing ratios (Sylgard 184 silicone oil: curing agent in a 10:1 ratio; Sylgard 527 part A: part B in a 1:1 ratio) were used. Each base blend was colored black with 1 w-% iron oxide powder. The production of base blends in the recommended mixing ratio ensured that complete cross-linking could take place for both components.

This is of crucial importance for the mechanical stability of the mixture of both PDMS types. Each base blend was degassed. Hereafter, the base blends were mixed with each other in a 1:1 ratio and a 1:5 ratio, respectively. Pure Sylgard 184, respectively the again degassed mixtures of Sylgard 184 and Sylgard 527, were poured into a mould for hemispheres. The hemisphere moulds had a diameter of 12 mm. The casted hemispheres were hardened for 12 h at 60 °C. For at least another 36 h, they were cured at room temperature. Two hemispheres were bonded together with a thin film of the corresponding newly produced PDMS compound. The bonded spheres were cured for at least another 48 h at room temperature.

In the standard case, experiments were performed with the spheres in the dry state after fabrication. For comparison, spheres that had remained in silicone oil after the first experiment showed swelling effects. It was observed that size, weight and consequently the density of the sphere changed over time. This suggests that the spheres swell by diffusion of the silicone oil into the elastomer-gel matrix. However, a significant change in the material properties was only observed after the spheres had been immersed in the oil for at least two days. Consequently, swelling effects are not to be expected during the short period of the experiment. A defined initial state could be ensured for the experiments.

To compare soft PDMS spheres with rigid spheres whose density is in a range comparable to that of PDMS, spheres made of epoxy resin were fabricated. Blends of Casting Resin MS 1000 by Weicon® were mixed according to the manufacturer's recommendations and colored black with 1 w-% iron oxide powder. The same hemisphere moulds as for the soft spheres were used to cast the hemispheres. The hemispheres were hardened according to the manufacturer's recommendation. The following steps were analogous to the fabrication of the soft spheres.

### 3.3. Material properties

The solid density of the spheres was determined directly after the sedimentation experiments. This ensured to be able to correctly represent the current state of the density. The solid density was determined using a hydrostatic balance (Sartorius® ENTRIS BCE 224i-1s analytical balance + density determination kit YDK03 for determination of solid and liquid densities). The balance has a readability of 0.1 mg. Density measurements were performed in distilled water. The density of silicone oil was also determined hydrostatically at room temperature. The corresponding material data for the solid and liquid density can be found in table 1.

The dynamic viscosity $\eta$ of the silicone oil under experimental conditions was measured to be about 980 mPa·s using a rotational viscometer (IKA® Rotavisc lo-vi).

The elastic material properties were determined experimentally by means of compression tests. Cylindrical specimens of the PDMS mixtures also used in the model experiments were fabricated. The compression tests were in accordance with DIN ISO 7743:2016-08. A Schenk Trebel RSA25 universal testing machine with a piezoelectric load cell and potentiometric displacement sensor was used to record the force-displacement measurement data in a displacement-controlled test ($v_{\Delta l} = 10$ mm·min$^{-1}$). To avoid friction effects, silicone oil with a viscosity of 1000 cSt was used as lubricant. The Young's elastic modulus of each specimen was determined from the slope of the compressive stress-strain curve in the linear-elastic range of small deformations. The experimentally determined averaged Young's moduli are shown in table 1.

| Material Properties | | | |
|---|---|---|---|
| **Liquid Density** | $\rho$ | $971.28 \pm 0.134$ | kg·m$^{-3}$ |
| **Liquid Dynamic Viscosity** | $\eta$ | $0.9797 \pm 0.012$ | Pa·s |
| **Solid Densities** | | | |
| Epoxy resin | $\rho_{\text{Epoxy}}$ | $1159.21 \pm 1.17$ | kg·m$^{-3}$ |
| Sylgard 184 | $\rho_{\text{SYL184}}$ | $1040.01 \pm 1.90$ | kg·m$^{-3}$ |
| Sylgard 184:527 1:1 | $\rho_{\text{MIX11}}$ | $1004.75 \pm 1.32$ | kg·m$^{-3}$ |
| Sylgard 184:527 1:5 | $\rho_{\text{MIX15}}$ | $987.59 \pm 1.90$ | kg·m$^{-3}$ |
| **Young´s Moduli** | | | |
| Epoxy resin | $E_{\text{Epoxy}}$ | $2.9^{1)}$ | GPa |
| Sylgard 184 | $E_{\text{SYL184}}$ | $1712 \pm 82.55$ | kPa |
| Sylgard 184:527 1:1 | $E_{\text{MIX11}}$ | $936 \pm 33.44$ | kPa |
| Sylgard 184:527 1:5 | $E_{\text{MIX15}}$ | $135 \pm 13.14$ | kPa |

[1)] according to manufacturer information

Table 1: Measured material properties.

## 4.   Results and Discussion

### 4.1.  Velocity curves

#### 4.1.1. Rigid spheres

Results for the velocity measurements of the nonheavy rigid spheres made of epoxy resin show typical velocity curves for rigid particles under wall influence, see figure 3. As expected, sedimentation shows two phases. After the initial, transient phase the rigid spheres reach the terminal velocity, the second phase of sedimentation dynamics.

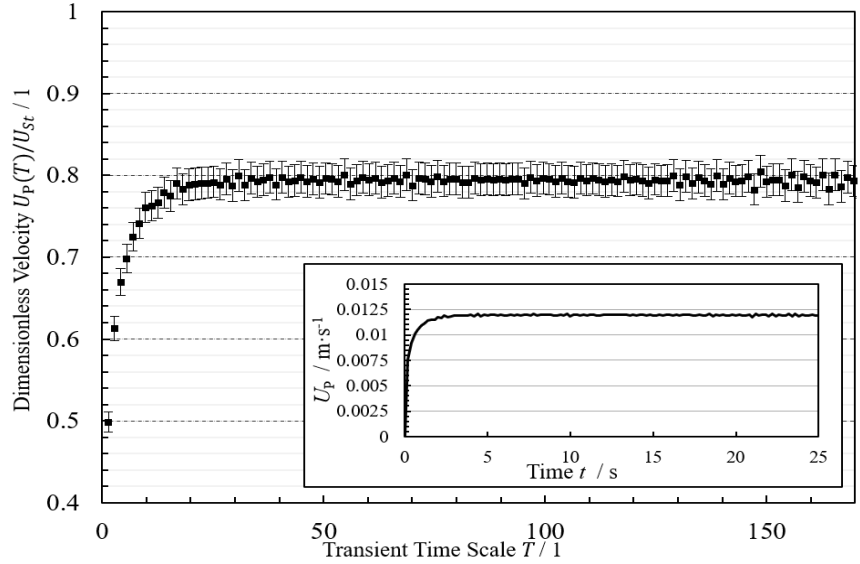

Figure 3: Dimensionless velocity plotted over the transient time scale of a sphere which starts sedimentation from rest. The sphere was made of epoxy resin with $E_\blacksquare = 2.9$ GPa, $\rho_{P,\blacksquare} = 1159.21 \pm 1.17$ kg $\cdot$ m$^{-3}$, $R_\blacksquare = 6 \cdot 10^{-3}$ m, $Re_{P,\blacksquare} = 0.179$, $\gamma_\blacksquare = 1.19$; Insert: absolute values of the measured particle velocity plotted over time.

The results are presented as absolute $U_P - t$ diagram and as dimensionless representation. As will be shown, the experiments with varying elasticity are not directly comparable on the dimensional time scale. The difference in density between materials and the mixtures lead to much faster sedimentation for the rigid spheres compared to the elastic spheres. Anyway, the dimensional quantities are shown for the sake of completeness.

The nondimensionalization of the velocity is performed by $U_{St}$ from equation (2) as characteristic scale. The dimensionless velocity is then plotted over the time scale resulting from the viscous diffusion time. This dimensionless time is also known as transient time scale $T$ (Tashibana and Kitasho 1976)

$$T = \frac{\eta t}{4\rho R^2} \ . \tag{5}$$

The measured terminal velocity of the rigid sphere is $\bar{U}_{P,term,\blacksquare} = 11.9$ mm $\cdot$ s$^{-1}$. The terminal Stokes velocity resulting from the measured material properties after equation (2) is $U_{St,\blacksquare} = 15.0 \pm 0.4$ mm $\cdot$ s$^{-1}$. Hence, the measured terminal velocity in the experiment is reduced to about 80 % of the velocity in an unbounded fluid. This is due to wall effects stemming from the four container walls, even though the particle sediments more than 11 times its radius away from the walls. This reduction is in accordance with existing literature data based on calculations. For example, according to Swan and Brady´s numerical calculations, a particle would reach a terminal velocity of approximately 85-90 % of the Stokes velocity when the sphere sediments at a distance of $d \approx 11R$ between two infinite walls. (Swan and Brady 2010) Since in their calculations just two parallel walls were assumed, a reduction to a value in the measured range of our experiments with four walls is realistic. Specifically with respect to four walls, a good agreement of the experimental values is obtained with equation (3) stemming from the increase in drag. With the correction factor found by Hensley and Papavassiliou in CFD simulations, a dimensionless velocity of $U_{P,corr}/U_{St} = 0.82$ is expected for the width of the container of 140 mm and a radius of 6 mm. (Hensley and Papavassiliou 2014)

Because the investigated particles classify as nonheavy, the transient phase is more extended than it would be the case for the sedimentation of particles with high particle-to-fluid-density ratio $\gamma \gg 1$. Solving the differential equation (1) with respect to $U_P$ neglecting the Basset history force, the theoretical terminal velocity for the epoxy resin spheres would be reached after about 0.1 seconds. In contrast, the experimental determined terminal velocity is reached after a time of about 3 seconds during the sphere sediments about $30\,\text{mm} = 5R$. Correspondingly, in the present case, initial effects like the added mass force and the Basset history force must be considered to describe the complete sedimentation dynamics adequately.

### 4.1.2. Elastic, deformable spheres

Velocity curves of elastic, deformable spheres differ the more from those of the rigid spheres, the lower the Young´s modulus $E$ becomes, see figure 4 *(a)-(c)*. In the following, we divide the sedimentation of the elastic spheres in four phases (figure 4*(d)*). The phases are: I. the transient acceleration and II. the first plateau. The dimensionless velocity on the first plateau is comparable with the dimensionless velocity of the rigid spheres as will be discussed in chapter 4.2. The rigid plateau is followed by phase III., the second acceleration. Ultimately, the elastic spheres reach phase IV., a second plateau on which the elastic spheres reach their terminal sedimentation velocity.

While the differences in the sedimentation of rigid and elastic spheres are not yet obvious at first glance for the spheres with the highest Young´s modulus of $E_{\text{SYL184}} = 1712\,\text{kPa}$ (figure 4 *(a)*), a reduction of the Young's modulus by about half ($E_{\text{MIX11}} = 936\,\text{kPa}$, figure 4 *(b)*) already leads to a clearer visibility of the phases. The four phases of sedimentation are most evident with the lowest Young´s modulus of $E_{\text{MIX15}} = 135 \pm 13.14\,\text{kPa}$ (figure 4 *(c)*). For this sphere, no clear steady-state condition is established over the measured distance. Based on the last measured values from $T \approx 1500$, a terminal velocity of $\overline{U}_{\text{P,term},\bullet} \approx 1.2\,\text{mm}\cdot\text{s}^{-1}$ can be estimated. All measured averaged terminal velocities and calculated Stokes's terminal velocities can be found in table 2.

| $E/\text{kPa}$ | $\overline{U}_{\text{P,term}}/\text{mm}\cdot\text{s}^{-1}$ | $U_{\text{St}}/\text{mm}\cdot\text{s}^{-1}$ |
|---|---|---|
| $1712 \pm 82.55$ | $4.5 \pm 0.0$ | $5.4 \pm 0.2$ |
| $936 \pm 33.44$ | $2.2 \pm 0.0$ | $2.7 \pm 0.1$ |
| $135 \pm 13.14$ | $1.2 \pm 0.0$ | $1.3 \pm 0.04$ |

Table 2: Velocity measurements

The error bars shown in figures 3 and 4 were calculated by means of systematic and statistical measurement errors. Systematic errors in our measurements include, for example, measurement inaccuracies of the measuring instruments or errors when scaling the data from image velocimetry (pixels to real length units). Statistical measurement errors include the deviations within multiple measurements of a measured quantity. Here, the error analysis was carried out just for the terminal Stokes velocity. The representation of the uncertainty for the time scale $T$ was omitted at this point for the sake of clarity.

Furthermore, only single experiments are presented in the figures. Experiments were performed with different specimens fabricated from each material and at least two measurements were performed for each specimen. It was shown that the experiments are consistent and reproducible in terms of quality and quantity within the scope of the deviations described above.

It should also be noted that the starting point for velocity measurement is the first image where the sphere is completely detached from the vacuum pipette. This inevitably leads to the fact that the time $T = 0$ may differ between the individual experiments (depending on the time step size selected for the evaluation).

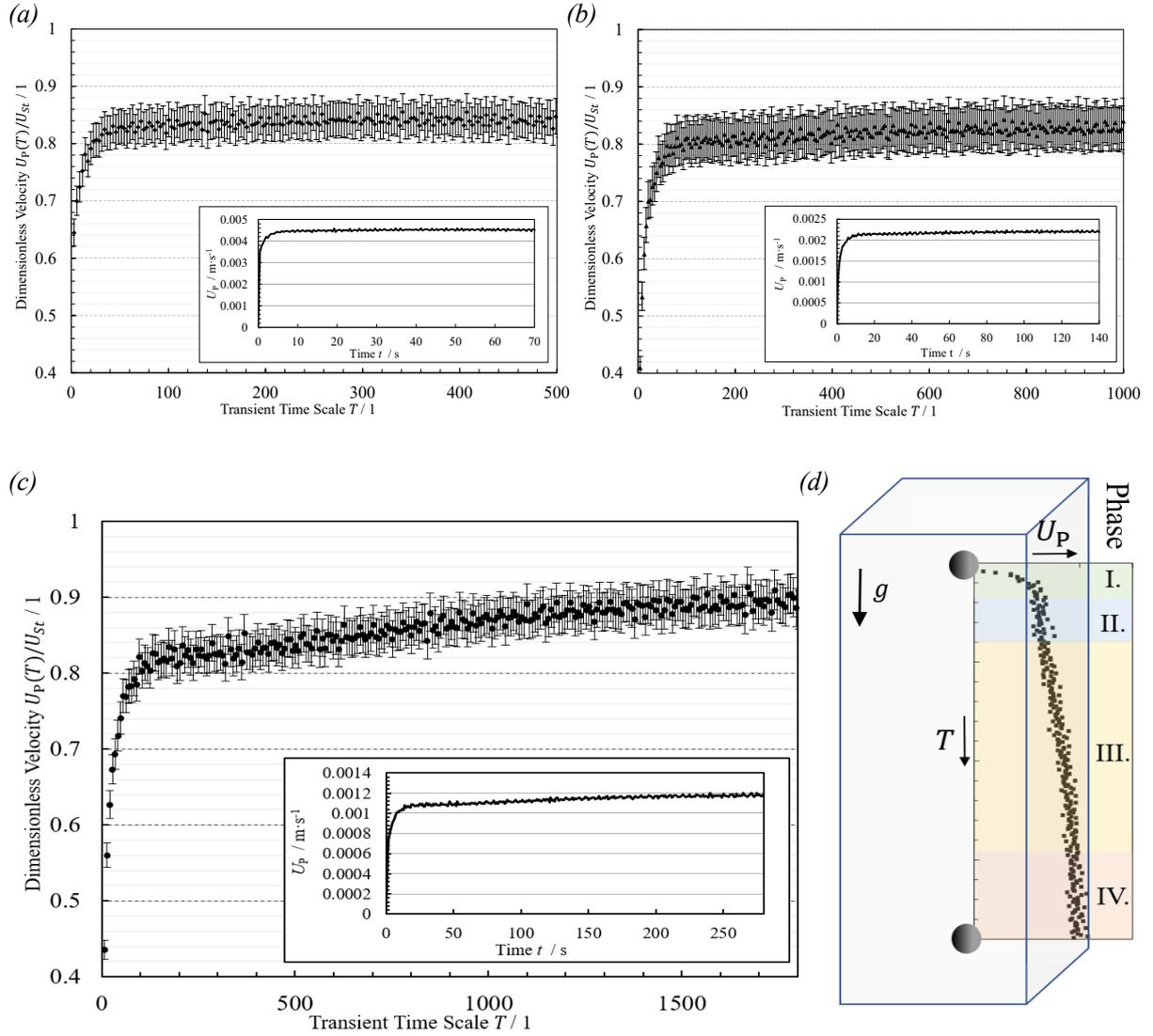

Figure 4: Dimensionless velocity plotted over the transient time scale of a sphere which starts sedimentation from rest made of *(a)* Sylgard 184 with $E_\blacklozenge = 1712 \pm 82.55$ kPa, $\rho_{P,\blacklozenge} = 1040.01 \pm 1.9$ kg $\cdot$ m$^{-3}$, $R_\blacklozenge = 0.00593$ m, $Re_{P,\blacklozenge} = 0.061$, $\gamma_\blacklozenge = 1.07$; *(b)* Sylgard 184:Sylgard 527 in a 1:1 mixing ratio with $E_\blacktriangle = 936 \pm 33.44$ kPa, $\rho_{P,\blacktriangle} = 1004.75 \pm 1.32$ kg $\cdot$ m$^{-3}$, $R_\blacktriangle = 0.00598$ m, $Re_{P,\blacktriangle} = 0.032$ $\gamma_\blacktriangle = 1.03$; *(c)* Sylgard 184:Sylgard 527 in a 1:5 mixing ratio with $E_\bullet = 135 \pm 13.14$ kPa, $\rho_{P,\bullet} = 987.59 \pm 1.9$ kg $\cdot$ m$^{-3}$, $R_\bullet = 0.006$ m, $Re_{P,\bullet} = 0.016$, $\gamma_\bullet = 1.02$; Inserts *(a)-(c)*: absolute values of the measured particle velocities plotted over time; *(d)* schematic of elastic particle sedimentation showing the four phases of sedimentation

### 4.2. Phase I. & II.: Rigid plateau and elastic time scale

When the dynamics of rigid and elastic spheres are compared, similarities arise during the initial phases. First, all soft model particles presumably reach a first plateau after a first, transient acceleration phase. This plateau lies at a value between ~0.8 and ~0.83 considering the mean value of the dimensionless velocity in this range. This velocity is just the same as the terminal velocity of the rigid spheres within the uncertainty of the measurements. It is hence reasonable to assume that all particles – both rigid and soft – initially reach the same reduced stationary state, which is only due to the wall influence. For a better overview in the following diagrams, we therefore slightly shift all curves within the measurement uncertainty to a reference value of 0.81 within this range.

However, the velocity-time behavior of the different spheres is not directly comparable to each other due to the different density ratios following from the individual mixing ratios (table 1) needed to fabricate spheres of varying elasticity. Consequently, inertial influences are different for each type of sphere, see figure 5 *(a)*.

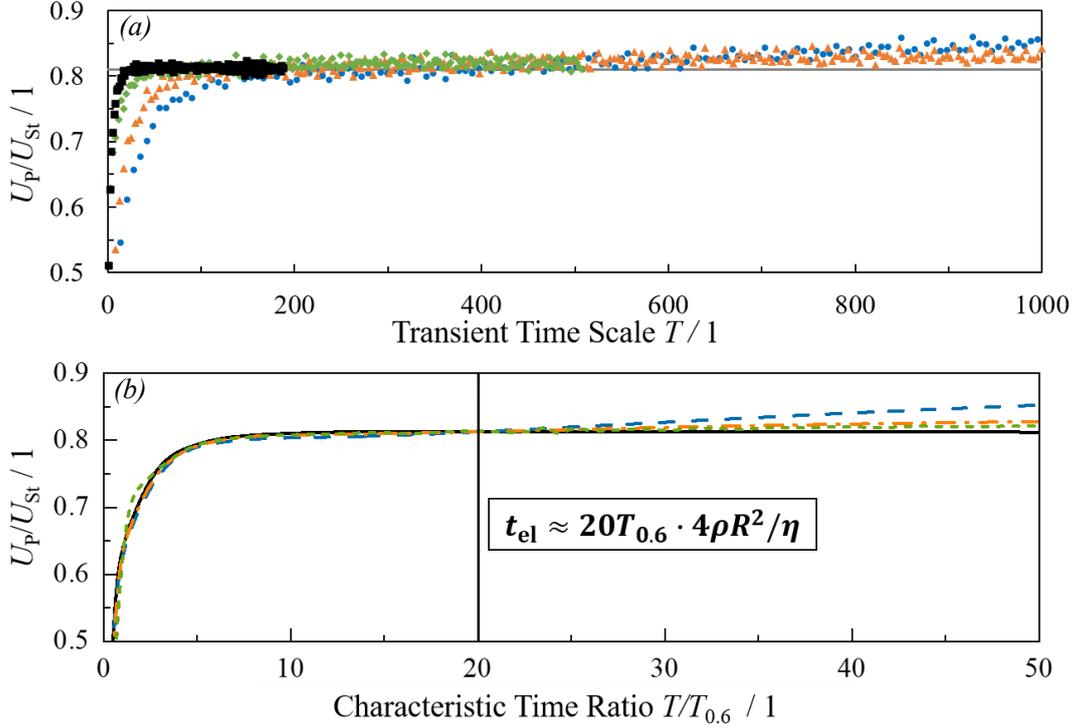

Figure 5: *(a)* Dimensionless velocities of the experiments from figures 3 and 4 (■ epoxy resin, ♦ Sylgard 184, ▲ Mix 1:1, ● Mix 1:5) fitted to a value of 0.81 at the first plateau and plotted against the transient time scale; *(b)* Interpolated dimensionless velocities of the experiments from figures 3 and 4 (— epoxy resin, - - -Sylgard 184, — · —Mix 1:1, — — — Mix 1:5) fitted to a value of 0.81 at the first plateau and plotted against the characteristic time ratio $T/T_{0.6}$. The interpolation was performed with cubic B-splines.

To make the experimental data of all experiments comparable on the time scale, the time delay of the increase due to inertia must be eliminated. One way to do this is reported by R.B. Moorman in 1955 in his dissertation (Moorman 1955). If the transient range is resolved, an inflection point of the velocity-time curve appears. This point marks the time from which the frictional forces become more significant and ultimately more dominant than the inertial forces in the transient range. Experimentally determining this inflection point to be located at about 60% of the terminal velocity, i.e. $U_P/U_{St} \approx 0.6$ at the corresponding time $T_{0.6}$, R.B. Moorman could map similar curves from the sedimentation of rigid spheres on a universal curve upon scaling $T$ with $T_{0.6}$. (Tashibana and Kitasho 1976) Following the same procedure, also the sedimentation velocities for elastic spheres fall on top of each other within the first two phases (figure 5 *(b)*). This supports the assumption that at the beginning of the sedimentation, elastic spheres follow the same principles as rigid spheres: inertial and frictional forces determine the transient acceleration until a steady state is reached. For the investigated spheres, any elastic effects obviously kick in at a time scale that is larger than the time scale for establishing this first plateau.

In the experiments, after reaching the first plateau, all curves start to rise again from a characteristic time ratio of $T/T_{0.6} \approx 20$ (figure 5 *(b)*). This allows an estimation and calculation of a time $t_{el}$ with

$$t_{el} \approx \frac{20 T_{0.6} \cdot 4 \rho R^2}{\eta} \ . \tag{6}$$

From this time on, an elastic effect leads to a significant change in the particle dynamics and to an increase in the sedimentation velocity due to elasticity. Furthermore, $t_{el}$ marks the end of the second sedimentation phase.

It should be mentioned that the normalization after Moorman in figure 5 was performed with respect to the sedimentation velocity in an unbounded fluid ($U_P/U_{St} \approx 0.6$). It remains to be clarified whether in the case of wall influence and consequently lower velocities, the inflection point shifts accordingly, here to a point for $T$ at $U_P/U_{St} \approx 0.6 * 0.81$. Experimentally, the consequences for $T_{0.6}$ are not large due to the steep increase in velocity. An experimental investigation of this inflection point would however require a large temporal resolution of the transient phase and is a potential field for further investigations. In the present case, the chosen approach is viable insofar as a mapping of the curves on top of each other could be established.

### 4.3. Phase III.: Second acceleration

As described above, a second, elastic acceleration starts about the characteristic time $t_{el}$. Alternatively, the sedimentation velocity can be considered as a function of position in the direction of gravitational acceleration (figure 6). All elastic spheres sediment with a velocity compareable to the rigid sphere´s velocity until the position $x_{el} \approx 10 \cdot R$ is reached. At this characteristic elastic onset position, the second acceleration begins. The extension of the second acceleration, i.e. the distance $\Delta x_{el}$ between the characteristic onset $x_{el}$ and the second plateau with the final velocity $\overline{U}_{P,\text{term}}$, depends on the elasticity of the sphere. The more deformable ($E \downarrow$) the sphere is, the longer is the distance $\Delta x_{el}$ until the stationary state is reached. For the present data, the length of the second acceleration phase due to elasticity can be approximately described in terms of the sphere radius $R$ as function of the Young´s Modulus $E$ (in Pa)

$$\frac{\Delta x_{el}}{R} \approx -\frac{E}{\text{Pa}} \cdot 10^{-5} + 36 \,. \tag{7}$$

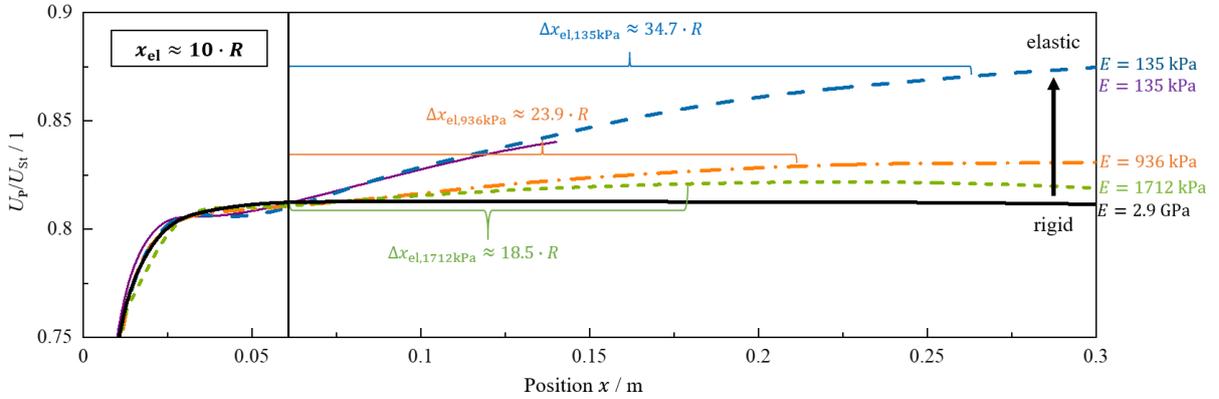

**Figure 6:** Interpolated dimensionless velocities of the experiments from figures 3 and 4 (— epoxy resin, - - -Sylgard 184, – · – · –Mix 1:1, – – – Mix 1:5, — Mix 1:5 deep immersed) fitted to a value of 0.81 at the first plateau and plotted against the Position $x$ – rigid sedimentation behavior is found until position $x_{el} \approx 10 \cdot R$ is reached.

As already described in section 4.1.2. with respect to time, the sphere with the lowest Young´s modulus seems to reach the terminal velocity only at the last measured positions. With respect to the three phases of sedimentation discussed so far, the elastic sphere sedimented nearly $45 \cdot R$ until the terminal velocity of the second plateau was reached. Thus, it becomes clear that deformable particles do not sediment stationary over long distances and show a long-term unsteady behavior.

To determine the relationship between Young´s modulus and the distance $\Delta x_{el}$ in a more universal fashion, further experiments are needed. For example, it should be clarified under which conditions a linear relationship as in equation (7) would hold.

To check any possible influence of hydrostatic pressure on the velocity behavior, experiments with a larger immersion depth were performed (violet curve in figure 6). The same sphere with a Young´s modulus of 135 kPa was immersed about 45 mm deeper before release. The deeper immersed sphere showed the same velocity behavior as the less immersed sphere. Consequently, no influence of the hydrostatic pressure (at least not at this immersion depths) could be observed.

Under the investigated conditions, the sedimentation dynamics of elastic particles could be well described in four phases. Both the characteristic elastic time $t_{el}$ and the characteristic onset position $x_{el}$ show a dependence on the particle radius. Especially in the case of small and highly deformable particles, it could be possible that elastic effects would kick in earlier. Then it would no longer be possible to distinguish between the acceleration phases and effectively two phases would be observed. However, the resulting overall acceleration phase would still be much longer than in the case of comparable rigid particles.

### 4.4. Phase IV.: Second plateau - elastic terminal velocity

After the second acceleration, the fourth phase of sedimentation of elastic spheres starts. In this final phase, the elastic spheres reach the terminal sedimentation velocity. The more deformable the spheres are, the faster are the spheres and the higher is the measured increase in velocity relative to the first plateau (see figure 6). The measured relative increase to the rigid sedimentation velocity $\Delta_{el,exp}$ reaches up to nearly 9 % for the softest spheres (exact values see data table in figure 7).

To our knowledge, no equation exists decribing the increase in velocity of elastic spheres in a bounded fluid. There is merely the expression proposed by Murata for sedimentation in an unbounded fluid (equation (4)). It predicts an increase to the terminal Stokes velocity of $\Delta_{el,M}$. (Murata 1980) Hence, the increase in sedimentation velocity obtained experimentally is not directly comparable to the sedimentation velocity for elastic spheres proposed by Murata in equation (4). However, the comparison of Murata´s predicted increase and the experimental obtained increase can be used to shed light on possible additional wall influences on the increase in velocity.

The increase according to Murata $\Delta_{el,M}$ can be determined using known material parameters. No exact measured values for the Poisson´s ratio of the PDMS mixtures are available. Therefore, the increase according to Murata was calculated for various Poisson´s ratios. Based on measurements for pure Sylgard 184, it is assumed that the Poisson's ratio of all mixtures has a value between $\nu = 0.490$ and $\nu = 0.499$. (Müller et al. 2019) The percentage increase from the experimental data is calculated from the mean values of the velocities of the first plateau and the second plateau.

Comparing the calculated increase according to Murata´s equation to the measured increase in the experiments, shows that the values fundamentally differ in the order of magnitude, but slightly correlate in the prefactor (see data table figure 7). The measured increase in velocity relative to the first plateau is up to an order of $\mathcal{O}(10^{10}$ %$)$ higher than the percentage increase predicted by Murata. In total, the increase according to Muarata $\Delta_{el,M}$ in an unbounded fluid is almost negligible for Poisson´s ratios in the previously mentioned range. In contrast, the measured increase of nearly 9% is not negligible, particularly with regard to applications. Exemplarily, the data table in figure 7 shows the percentage increases $\Delta_{el,M}$ for a Poisson's ratio of 0.491 assumed for all PDMS mixtures.

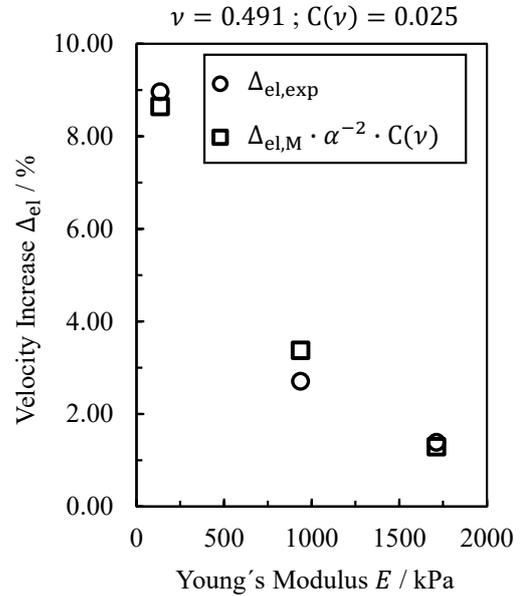

| $E$ / kPa | $\Delta_{el,exp}$ / % | $\Delta_{el,M}$ / % | $0.025 \cdot \Delta_{el,M} \cdot \alpha^{-2}$ /% |
|---|---|---|---|
| 1712 | 1.39 | $1.24\cdot10^{-10}$ | 1.30 |
| 936 | 2.71 | $2.60\cdot10^{-10}$ | 3.37 |
| 135 | 8.96 | $7.65\cdot10^{-9}$ | 8.64 |

**Figure 7:** Data table for experimental values of $\Delta_{el,exp}$ and theoretical values of $\Delta_{el,M}$ calculated with a Poisson´s ratio of $\nu = 0.491$ (left); Experimental determined and by factor $\alpha^{-2} \cdot C(\nu)$ fitted values of the velocity increase to the rigid velocity plotted against the Young´s modulus $E$ (right).

Although the values of $\Delta_{el}$ do not fit to each other at the first glance, a closer look on the measured increase compared to the predicted increase shows that they just seem to differ by a factor of the perturpation parameter $\alpha^2$ and some constant $C(\nu)$ which depends on the Poisson´s ratio (figure 7 (right)).

$$\Delta_{el,exp} \approx C(\nu) \cdot \alpha^{-2} \cdot \Delta_{el,M} \qquad (8)$$

More investigations on this relation are needed. Calculation of the specific constants $K_0$ and $K_1$ from Murata's equation suggests that the sphere deforms into a prolate ellipsoid while sedimentation. However, no apparent deformation could be detected with our measurement method. The deformation must be small, which is also reflected in the calculated parameter $\alpha$ (viscous stresses vs. mechanical stresses), which is always $\ll 1$. It is unclear at this point whether the walls are causal for the much higher increase to the rigid velocity compared to Murata´s calculated increase. It is conceivable that wall effects have a much larger influence on the deformation than the pure elastohydrodynamic effect during sedimentation.

In order to further investigate the effect of walls on the terminal sedimentation velocity, more experiments are needed, especially with varying radius to container size ratios. Experiments for the investigation of sedimentation dynamics in the unbounded fluid are challenging because it must be ensured that the wall distances are sufficiently large to exclude wall influences. This would either lead to significantly larger container dimensions and thus problems in the implementation of such an experiment. Or, if smaller spheres are chosen while keeping the container size constant, the spheres are not heavy enough to overcome cohesion to the pipette. This was observed with spheres of the 1:5 mixture of Sylgard 184 and Sylgard 527 that have a radius of $R \approx 4$ mm. With reduction of weight and decrease of the Reynolds numbers, no results could be obtained so far in tests already carried out.

## 5. Conclusion

In this paper, we presented experimentally derived characteristics of sedimentation of elastic, deformable and nonheavy spheres in a rectangular duct. The experiments gave fundamental insights into the fluid-particle dynamics of elastic particles in bounded fluids and at low Reynolds numbers. We observed that the sedimentation of elastic spheres under the influence of walls differs fundamentally from that of rigid spheres.

Sedimentation of rigid and elastic spheres with a radius of 6 mm was performed in a rectangular duct at low Reynolds numbers. Sedimentation dynamics of rigid spheres was as expected. Velocities of ~81% of the Stokes velocity were measured for the rigid spheres. The reduction of the velocity is due to the influence of the four surrounding parallel walls in the rectangular duct. The measured reduction in velocity was expected and agrees with theoretical calculations from literature. The experiments with the elastic model particles showed a completely different velocity-time behavior, or velocity-displacement behavior respectively, than rigid particles. Elastic spheres showed four phases of sedimentation instead of two phases. The first phase showed a transient acceleration and the second phase a first velocity plateau that corresponds to the dimensionless velocity of the rigid spheres. Thereafter, for all elastic spheres a second acceleration began after a distance of about $10 \cdot R$. This characteristic onset position corresponds to a time of $t_{el} \approx 20T_{0.6} \cdot 4\rho R^2 \eta^{-1}$. From this characteristic time on, the elasticity has a significant influence on the sedimentation of elastic spheres in the bounded fluid. The third phase of sedimentation showed further acceleration that depends strongly on the Young's modulus. The more deformable the spheres were, the longer the second acceleration lasted and the further the distance was until the terminal velocity of the second plateau was reached. It was shown, that the softest spheres sedimented up to $45 \cdot R$ until the terminal velocity of the second plateau was reached. Furthermore, the measured percentage increase $\Delta_{el,exp}$ compared to the rigid velocity plateau was found to be up to 9 %. This long-term unsteady behavior and the much higher terminal velocities both are relevant for calculations, but they also have a significant impact on applications, e.g. process design, since these deviations from the known rigid particle behavior are not negligible.

With the help of our experimental data and the characteristic quantities obtained, it is already possible to make some easy-to-calculate statements about the expected terminal velocity of elastic (micro-)particles under wall influence. Our findings and their discussions regarding the gaps that still exist, also encourage a great deal of further research. Our findings can be used as a basis for following investigations in this area, which in our opinion, especially with respect to the associated applications, has been little studied experimentally but also theoretically in the past. In particular, we see a lot of potential in conducting further experiments with more variations of Young's modulus and of the ratio of the radius to the duct size. Even sedimentation of elastic spheres in an unbounded fluid remains to be investigated experimentally. An overaching goal for the future would be the modeling of the relationship between the elastic material parameters and the particle dynamics. With the help of analytical expressions of the forces acting on particles or the sedimentation velocity as a function of the Young's modulus, the computational effort, e.g. for CFD simulations including elastic particle dynamics, would be greatly reduced.

### Acknowledgements.

Funded by the Deutsche Forschungsgemeinschaft (DFG, German Research Foundation) – Project-ID 172116086 – SFB 926. The experiments were performed in frame of subproject A10 "Elastohydrodynamic interactions of particles flowing over microstructured surfaces" in the Collaborative Research Center 926 (CRC926 MICOS). Thanks to our CRC926 collaboration partners from the Working Group Materials Testing (AWP), Technische Universität Kaiserslautern, D-67663 Kaiserslautern, Germany under Prof. Eberhard Kerscher (subproject B04 of CRC926). Special thanks to Markus Burmeister (AWP) for material testing of the elastic specimens and for the editorial support.

## Declaration of Interests.

The authors report no conflict of interest.

## Appendix

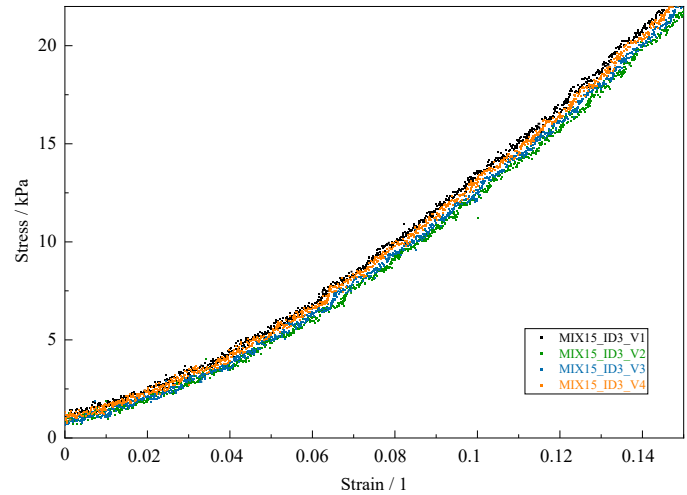

Figure A1: Stress-strain curve of four cycles (load and unloading) of a test with a cylinder fabricated from a Sylgard 184 and Sylgard 527 in a 1:5 mixing ratio lubricated with silicone oil; Since no standard bodies for testing were used and the recorded displacement of the sensor is not adjusted to the sample height, there are intercepts (curve does not start at point (0/0)). To determine Young's moduli and thereby exclude initial and frictional effects, in this example the slopes of in total 0.0.5 absolute strain in the range of small deformations are determined by a linear fit (here in the range [0.025;0.075])